\documentclass[reqno]{article}%12pt%
\usepackage{alltt}
\usepackage{graphicx}
\title{Objective quantum theory based on the CP(N-1) affine gauge fields}
\author{Peter Leifer}
\date{ Cathedra of Informatics, Crimea State Engineering and
Pedagogical University, \\
21 Sevastopolskaya st., 95015 Simferopol, Crimea, Ukraine }
\begin{document}
\maketitle
\begin{abstract}
The ordinary linear quantum theory predicts the quantum correlations
at any distance (the universal superposition principle). It creates
the decoherence problem since quantum interactions entangle states
into non-separable combination. On the other hand the linear quantum
theory prevents the existence of the localizable solutions, and
after all, leads to the divergences problem in the quantum field
theory. In order to overcome these difficulties the non-perturbative
nonlinearity originated by the curvature of the compact quantum
phase space has been used.
\end{abstract}


\section{Introduction}
Non-linearity in quantum theory has been invoked in order to build
the objective quantum theory and to prevent the unlimited spread out
of the observable fields by the gravitational self-potential
\cite{Jones1,Jones2}. But Newtonian quantum gravity in the present
form is not effective for the shaping wave-packets of elementary
particle size since the characteristic scale of the ground-state
wave-packet obtained from the gravitational Schr\"odinger equation
for nucleon masses is around $10^{23} m$ \cite{Jones2}.

There is a different group of works that emphasize the formulation
of {\it the standard quantum mechanics} in quantum phase space (QPS)
represented by the complex projective Hilbert space $CP(N-1)$
\cite{CMP,Hughston1,Hughston2,Ash,AdHr}. I think, however, that a
consistent and prolific theory based on such QPS should be connected
with serious deviations from the standard quantum scheme. Such a
modification must, of course, preserve all achievements of
de~Broglie-Heisenberg-Schr\"odinger-Dirac linear theory in a natural
way. One may think about attempts to establish a deductive approach
to the quantum theory.

Standard quantum mechanics (QM) treats electron as pointwise
particle but it is `wrapped' in so-called de~Broglie-Schr\"odinger
fields of probability. Quantum field theory (QFT) uses same
classical space-time coordinates of the pointwise particle as
`indices' whereas the fields are operators acting in some Hilbert
state space (frequently in Fock space). QM and QFT take account of
the non-commutative nature of the dynamical variables but the
interaction between pointwise particles and the relativistic
invariance are borrowed from the classical theory. These are the
sources of the singular functions involved in QFT. It is useful to
understand the true reason of these difficulties.

The special and general relativity are based on the possibility to
detect locally the coincidence of two pointwise events of different
nature. As such the ``state" of the local clock gives us local
coordinates - the ``state" of the incoming train \cite{Einstein1}.
In the classical case the notions of the ``clock" and the ``train"
are intuitively clear. Furthermore, Einstein especially notes that
he does not discuss the inaccuracy of the simultaneous of two {\it
approximately coincided events} that should be overcame by some
abstraction \cite{Einstein1}. This abstraction is of course the
neglect of finite sizes (and all internal degrees of freedom) of the
both real clock and train. It gives the representation of these
``states" by mathematical points in space-time. Thereby the local
identification of two events is the formal source of the classical
relativistic theory. But in the quantum case it is impossible since
the localization of quantum particles is state-dependent
\cite{W,Heg,A97}. Hence the identification of quantum events
(transitions) requires a physically motivated operational procedure
with corresponding mathematical description.

Therefore it is inconsistent to start the development of the quantum
theory from the space-time symmetries because just the space-time
properties should be established in some approximation to internal
quantum dynamics, i.e. literally {\it a posteriori}. Namely, the
quantum measurement with help of the ``quantum question" leads
locally to the Lorentz transformations of its spinor components,
and, on the other hand, to dynamical (state-dependent) space-time
coordinatization. That is, instead of the representation of the
Poincare group in some extended Hilbert space, I used an ``inverse
representation" of the $SU(N)$ by solutions of relativistic
quasi-linear PDE in the dynamical space-time. It is in fact the one
of the possible realizations of L. de Broglie idea about the ``wave
with a hump" \cite{dB}.

In the present article I propose a non-linear relativistic 4D field
model originated by the internal dynamics in QPS $CP(N-1)$
\cite{Le1,Le2}. This is the development of the ideas used in
\cite{Le3}. There is no initially distinction between `particle' and
`field', and the space-time manifold is derivable. Quantum
measurements will be described in terms of parallel transport of the
local dynamical variables.

\section{The Action Quantization}
Schr\"odinger sharply denied the existence of so-called ``quantum
jumps" during the process of emission/absorption of the quants of
energy (particles) \cite{Schr1,Schr2}. Leaving the question about
the nature of quantum particles outside of consideration, he thought
about these processes as a resonance of the de Broglie waves that
phenomenologically may look like ``jumps" between two ``energy
levels". The second quantization method formally avoids these
questions but there are at least two reasons for its modification:

{\it First.} In the second quantization method one has formally
given particles whose properties are defined by some commutation
relations between creation-annihilation operators. Note, that the
commutation relations are only the simplest consequence of the
curvature of the dynamical group manifold in the vicinity of the
group's unit (in algebra). Dynamical processes require, however,
finite group transformations and, hence, the global group structure.
The main my technical idea is to use vector fields over a group
manifold instead of Dirac's abstract q-numbers. This scheme
therefore seeks the dynamical nature of the creation and
annihilation processes of quantum particles.

{\it Second.} The quantum particles (energy bundles) should
gravitate. Hence, strictly speaking, their behavior cannot be
described as a linear superposition. Therefore the ordinary second
quantization method (creation-annihilation of free particles) is
merely a good approximate scheme due to the weakness of gravity.
Thereby the creation and annihilation of particles are time
consuming dynamical non-linear processes. So, linear operators of
creation and annihilation (in Dirac sense) do exist as approximate
quantities.

POSTULATE 1.

\noindent {\it There are elementary quantum states $|\hbar a>,
a=0,1,...$ belonging to the Fock space of an abstract Planck
oscillator whose states correspond to the quantum motions with given
number of Planck action quanta}.

 One may image some {\it ``elementary quantum states"
(EAS) $|\hbar a>$ as a quantum motions with entire number $a$ of the
action quanta}. These $a,b,c,...$ takes the place of the ``principle
quantum number" serving as discrete indices $0 \leq a,b,c... <~
\infty$. Since the action in itself does not create gravity, it is
possible to create the linear superposition of $|\hbar
a>=(a!)^{-1/2} ({\hat \eta^+})^a|\hbar 0>$ constituting $SU(\infty)$
multiplete of the Planck's action quanta operator $\hat{S}=\hbar
{\hat \eta^+} {\hat \eta}$ with the spectrum $S_a=\hbar a$ in the
separable Hilbert space $\cal{H}$. Therefore, we shall primarily
quantize the action, and not the energy. The relative (local) vacuum
of some problem is not necessarily the state with minimal energy, it
is a state with an extremal of some action functional.

Generally the coherent superposition
\begin{eqnarray}
|F>=\sum_{a=0}^{\infty} f^a| \hbar a>,
\end{eqnarray}
may represent of a ground state or a ``vacuum" of some quantum
system with the action operator
\begin{eqnarray}
\hat{S}=\hbar A({\hat \eta^+} {\hat \eta}).
\end{eqnarray}
Such vacuum is more general than so-called ``$\theta$-vacuum" of QCD
\begin{eqnarray}
|\theta>=\sum_{a=-\infty}^{\infty} e^{i\theta a}| \hbar a>.
\end{eqnarray}
It is clear that  $|F>=|\theta>$ if $f^0=1, \quad f^n=2 \cos (\theta
n)$ for $n=1,...$ and if $|\theta>$ is absolutely convergent
series. The ``winding number" $a$ has here different sense as it was
mentioned above.

The space-time representation of EAS's and their coherent
superposition is postponed on the dynamical stage as it is described
below. We shall construct non-linear field equations describing
energy (frequency) distribution between EAS's $|\hbar a>$, whose
soliton-like solution provides the quantization of the dynamical
variables. Presumably, the stationary processes are represented by
stable particles and quasi-stationary processes are represented by
unstable resonances.

The action functional
\begin{eqnarray}
S[|F>]=\frac{<F|\hat{S}|F>}{<F|F>},
\end{eqnarray}
has the eigen-value $S[|\hbar a>]=\hbar a$ on the eigen-vector
$|\hbar a>$ of the operator $\hbar A({\hat \eta^+} {\hat
\eta})=\hbar {\hat \eta^+} {\hat \eta}$. This deviates in general
from this value on superposed states $|F>$ and of course under a
different choice of $\hat{S}=\hbar A({\hat \eta^+} {\hat \eta}) \neq
\hbar {\hat \eta^+} {\hat \eta}$. In order to study the variation of
the action functional on superposed states one need more details of
geometry of their superposition.

In fact only finite, say, $N$ elementary quantum states (EQS's)
($|\hbar 0>, |\hbar 1>,...,|\hbar (N-1)>$) may be involved in the
coherent superposition $|F>$. Then $\cal{H}=C^N$ and the ray space
$CP(\infty)$ will be restricted to finite dimensional $CP(N-1)$.
Hereafter we will use the indices as follows: $0\leq a,b \leq N$,
and $1\leq i,k,m,n,s \leq N-1$. This superposition physically
corresponds to the complete amplitude of some quantum motion.
Sometimes it may be interpreted as a extremal of action functional
of some classical variational problem.

The global vacuum $|\hbar 0>$ corresponds to the zero number of
action quanta in the places of the Universe far enough from stars
with pseudo-Euclidean metric in accordance with the

POSTULATE 2.

\noindent {\it `Mach's quantum principle': the Universe generates
the omnipresent average self-consistent cosmic potential coinciding
with the fundamental constant $g_{00}=\Phi_U = c^2$.}

Matter exists in the motion with a finite number of the action
quanta. The mass of some quantum particle gives the rate of
variation of the Universe potential $\Phi_U=c^2$ in accordance with
the de~Broglie  frequency $\omega=\frac{mc^2}{\hbar}$. Therefore
omnipresent $\Phi_U=c^2$ serves as a ``spring" of the ``local
internal clock'' showing rather the state-dependent time $\tau$ than
the ``world time" of Newton-Stueckelberg-Horwitz-Piron \cite{H1}.

Dynamics of the global vacuum state $|\hbar 0>$ or the ground state
of some problem $|G>$ should be sensitive to mass/energy
distribution and, hence, to the rate of variation of the Universe
potential $\Phi_U=c^2=G_N\frac{M_U}{R_U}$. If one assumes that the
universal linear density
$\rho_U=\frac{M_U}{R_U}=\frac{c^2}{G_N}=1.349\times 10^{27}[kg/m]$
preserves in some wave process with a mass $m$ per one wave length
$\lambda=v_{ph}T=\frac{2\pi v_{ph}}{\omega}$, i.e.
$\frac{m}{\lambda}=\rho_U$ then using the universal Einstein-de
Broglie relation $\frac{m}{\omega}=\frac{\hbar }{c^2}$ it is easy to
prove  that $\omega=\frac{2\pi v_{ph}}{\lambda}=k
v_{ph}=\frac{mc^2}{\hbar}$, since $k=\frac{2\pi c^2}{mG_N}$ and
$v_{ph}=\frac{m^2G_N}{h}$. The group velocity of this process is
$v_{gr}=\frac{d\omega}{dk}=\frac{d(k v_{ph})}{dk}=v_{ph}=c$.
Therefore one has $m=\sqrt{\frac{hc}{G_N}}=m_P$ and
$\lambda=\lambda_C=\frac{h}{mc}=\sqrt{\frac{hG_N}{c^3}}=l_P$.

This wave process is necessarily realized on the Planck's scale
because we treated the global vacuum deformation only as a linear
one-dimension wave perturbation of $\Phi_U=c^2$ whereas the quantum
ground state moves in fact in the Hilbert space $\cal{H}$. One needs
to study the general law of motion (``deformation") of ray of the
action state in the $CP(N-1)$. The wave propagation attendant to
this ``deformation" serves  as ``inverse representation" of $SU(N)$
in the dynamical (state-dependent) space-time (see below).

Since any ray of the action amplitude has isotropy group
$H=U(1)\times U(N)$ only the coset transformations $G/H=SU(N)/S[U(1)
\times U(N-1)]=CP(N-1)$ effectively act in $\cal{H}$. Therefore the
ray representation of $SU(N)$ in $C^N$, in particular, the embedding
of $H$ and $G/H$ in $G$, is a state-dependent parametrization.
Hence, there is a diffeomorphism between the space of the rays
marked by the local coordinates in the map
 $U_j:\{|G>,|g^j| \neq 0 \}, j>0$
\begin{equation}
\pi^i_{(j)}=\cases{\frac{g^i}{g^j},&if $ 1 \leq i < j$ \cr
\frac{g^{i+1}}{g^j}&if $j \leq i < N-1$}
\end{equation}\label{coor}
and the group manifold of the coset transformations
$G/H=SU(N)/S[U(1) \times U(N-1)]=CP(N-1)$. This diffeomorphism is
provided by the coefficient functions $\Phi^i_{\alpha}$ of the local
generators (see below). The choice of the map $U_j$ means, that the
comparison of quantum amplitudes refers to the amplitude with the
action $\hbar j$. The breakdown of $SU(N)$ symmetry on each action
amplitude to the isotropy group $H=U(1)\times U(N-1)$ contracts the
full dynamics down to $CP(N-1)$. The physical interpretation of
these transformations is given by the

POSTULATE 3.

\noindent {\it The unitary transformations of the action amplitudes
may be identified with physical fields; i.e., transformations of the
form} $U(\tau)=\exp(i\Omega^{\alpha}\hat{\lambda}_{\alpha}\tau)$,
where the field functions $\Omega^{\alpha}$ are the parameters of
the adjoint representations of $SU(N)$. {\it The coset
transformation $G/H=SU(N)/S[U(1)\times U(N-1)]=CP(N-1)$ is the
quantum analog of a classical force; its action is equivalent to
some physically distinguishable variation of GCS in $CP(N-1)$}.

Thus the quantum dynamics in the $CP(N-1)$ is similar to general
relativity dynamics, where due to the equivalence principle, gravity
is locally non-distinguishable from an accelerated reference frame
\cite{Einstein2}. But in general relativity one has the distinction
(by definition) between gravity (curvature) and its `matter' source.
In quantum physics, however, all physical fields are `matter' and
variation of these fields leads to the variation of basis in the
state space.

\section{Non-linear treatment of the eigen-problem}
The quantum mechanics assumes the priority of the Hamiltonian given
by some classical model which henceforth should be ``quantized". It
is known that this procedure is ambiquous. In order to avoid the
ambiguity, I intend to use a {\it quantum state} itself and the
invariant conditions of its conservation and perturbation. These
invariant conditions are rooted into the global geometry of the
dynamical group manifold. Namely, the geometry of $G=SU(N)$, the
isotropy group $H=U(1)\times U(N-1)$ of the pure quantum state, and
the coset $G/H=SU(N)/U(1)\times U(N-1)$ geometry, play an essential
role in the quantum state evolution (the super-relativity principle
\cite{Le4}). The stationary states (some eigen-states of action
operator, i.e. the states of motion with the least action) may be
treated as {\it initial conditions} for GCS evolution. Particulary
they may represent a local minimum of energy (vacuum).

Let me assume that $\left\{|\hbar a>\right\}_0^{N-1}$ is the basis
in Hilbert space $\mathcal{H}$. Then a typical vector $|F> \in
\mathcal{H}$ may be represented as a superposition
$|F>=\sum_0^{N-1}f^a|\hbar a>$. The eigen-problem may be formulated
for some hermitian dynamical variable $\hat{D}$ on these typical
vectors $\hat{D}|F>=\lambda_D|F> $. This equation may be written in
components as follows: $\sum_0^{N-1}D^a_b f^b=\lambda_D f^a$, where
$D^a_b=<a|\hat{D}|b>$.
\begin{equation}
\hat{D}=\sum_{a,b \geq 0} <a|\hat{D}|b> \hat{P}_{ab}=\sum_{a,b \geq
0}D_{ab}\hat{P}_{ab}=\sum_{a,b \geq 0} F^{\alpha}_D
\hat{\lambda}_{\alpha,(ab)} \hat{P}_{ab},
\end{equation}
where $\hat{P}_{ab}$ is projector. In particular, the Hamiltonian
has similar representation with $F^{\alpha}_H=\hbar \Omega^{\alpha}$
\cite{Le5}.

One has the spectrum of $\lambda_D:
\left\{\lambda_0,...,\lambda_{N-1} \right\}$ from the equation
$Det(\hat{D}-\lambda_D \hat{E})=0$ , and then one has the set of
equations $\hat{D}|D_p>=\lambda_p|D_p> $, where $p=0,...,N-1$ and
$|D_p>=\sum_0^{N-1}g^a_p|\hbar a>$ are eigen-vectors. It is worse
while to note here that the solution of this problem gives rather
rays than vectors, since eigen-vectors are defined up to the complex
factor. In other words we deal with rays or points of the non-linear
complex projective space $CP(N-1)$ for $N \times N$ matrix of the
linear operator acting on $C^N$. The Hilbert spaces of the infinite
dimension will be discussed later.

For each eigen-vector $ |D_p> $ corresponding $\lambda_p$ it is
possible to chose at least one such component $g^j_p$ of the
$|D_p>$, that $|g^j_p| \neq 0$. This choice defines in fact the map
$U_{j(p)}$ of the local projective coordinates for each
eigen-vecrtor
\begin{equation}
\pi^i_{j(p)}=\cases{\frac{g^i_p}{g^j_p},&if $ 1 \leq i < j$ \cr
\frac{g^{i+1}_p}{g^j_p}&if $j \leq i < N-1$}
\end{equation}\label{coor}
of the ray corresponding $|D_p>$ in $CP(N-1)$. Note, if all
$\pi^i_{j(p)}=0$ it means that one has the ``pure" state
$|D_p>=g^j_p|j>$ (without summation in $j$). Any different points of
the $CP(N-1)$ corresponds to the GCS's. They will be treated as
self-rays of some deformed action operator. Beside this I will treat
the superposition state $|G>=\sum_{a=0}^{N-1} g^a|a\hbar>$ as
``analytic continuation" of the of eigen-vector for an arbitrary set
of the local coordinates.

People frequently omit the index $p$, assuming that
$\lambda:=\lambda_p$, for $j=0$. Then they have, say, for the $N
\times N$ Hamiltonian matrix $\hat{H}$ the eigen-problem
$(\hat{H}-\lambda)|\psi>=0$ where I put $\psi^a:=g^a_0$.

In accordance with our assumption the $\lambda$ is such that $\psi^0
\neq 0$. Let then divide all equations by $\psi^0$. Introducing
local coordinates $\pi^i=\frac{\psi^i}{\psi^0}$, we get the system
of the non-homogeneous equations
\begin{eqnarray}
(H_{11}-\lambda )\pi^1 +...+ H_{1 i}\pi^i +...+ H_{1 N-1}\pi^{N-1}
=- H_{10}  \cr H_{21}\pi^1 + (H_{22}-\lambda)\pi^2 +...+ H_{2
i}\pi^i +...+ H_{2 N-1}\pi^{N-1} =- H_{20} \cr . \cr . \cr . \cr
H_{N-1 1}\pi^1 +...+ H_{N-1 i}\pi^i +...+ (H_{N-1 N-1} - \lambda)
\pi^{N-1} = -H_{N-1 0},
\end{eqnarray}
where the first equation
\begin{eqnarray}
H_{01}\pi^1 +...+ H_{0i}\pi^i +...+ H_{0 N-1}\pi^{N-1}
=-(H_{00}-\lambda)
\end{eqnarray}
is omitted. If $D=det(H_{ik}-\lambda \delta_{ik}) \neq 0, i \neq 0,
k \neq 0$ then the single defined solutions of this system may be
expressed through the Cramer's rule
\begin{eqnarray}
\pi^1=\frac{D_1}{D},...,\pi^{N-1}=\frac{D_{N-1}}{D}.
\end{eqnarray}
It is easy to see that these solutions being substituted into the
first omitted equation give us simply re-formulated initial
characteristic equation of the eigen-problem. Therefore one has the
single valued ray solution of the eigen-problem expressed in local
coordinates instead of the vector solution with additional freedom
of a complex scale multiplication.

This approach does not give essential advantage for a single
operator and it only shows that the formulation in local coordinates
is quite natural. But if one tries to understand how the
multi-dimensional variation of the hermitian operator included in a
parameterized family, the local formulation is inevitable. First of
all it is interesting to know the invariants of such variations. In
particular, the quantum measurement of dynamical variable
represented by hermitian $N \times N$ matrix should be described in
the spirit of typical polarization measurement of the coherent
photons \cite{Le6}. I will put below the sketch that depicts the
modulation measurement of the photons polarization state during the
operational ``travel" on the Poincar\'e sphere.

The initial state of the coherent photons $|x>$ is modulated passing
through an optically active medium (say using the Faraday effect in
YIG film magnetized along the main axes in the $z$-direction by a
harmonic magnetic field with frequency $\Omega$ and the angle
amplitude $\beta$). Formally this process may be described by the
action of the unitary matrix ${\hat h}_{os_3}$ belonging to the
isotropy group of $|R>$ \cite{Le3}. Then the coherence vector will
oscillate along the equator of the Poincar\'e sphere. The next step
is the dragging of the oscillating state $|x'(t)>=\hat{h_{os_3}}|x>$
with frequency $\omega$ up to the ``north pole'' corresponding to
the state $|R>$. In fact this is the motion of the coherence vector.
This may be achieved by the variation of the azimuth of the linear
polarized state from $\frac{\theta}{2}=-\frac{\pi}{4}$ up to
$\frac{\theta}{2}=\frac{\pi}{4}$ with help of the dense flint of
appropriate length embedded into the sweeping magnetic field.
Further this beam should pass the $\lambda /4$ plate. This process
of variation of the ellipticity of the polarization ellipse may be
described by the unitary matrix ${\hat b}_{os'_1}$ belonging to the
coset homogeneous sub-manifold $U(2)/[U(1) \times U(1)]=CP(1)$ of
the dynamical group $U(2)$ \cite{Le6}. This dragging without
modulation leads to the evolution of the initial state along the
geodesic of $CP(1)$ and the trace of the coherent vector is the
meridian of the Poincar\'e sphere between the equator and one of the
poles. The modulation deforms both the geodesic and the
corresponding trace of the coherence vector on the Poincar\'e sphere
during such unitary evolution.

The action of the $\lambda /4$ plate depends upon the state of the
incoming beam (the relative orientation of the fast axes of the
plate and the polarization of the beam). Furthermore, only relative
phases and amplitudes of photons in the beam have a physical meaning
for the $\lambda /4$ plate. Neither the absolute amplitude
(intensity of the beam), nor the general phase affect the
polarization character of the outgoing state. It means that the
device action depends only upon the local coordinates $\pi^1
=\frac{\Psi^1}{\Psi^0} \in CP(1)$. Small relative re-orientation of
the $\lambda /4$ plate and the incoming beam leads to a small
variation of the outgoing state. This means that the $\lambda /4$
plate re-orientation generates the tangent vector to $CP(1)$. It is
natural to discuss the two components of such a vector: velocities
of the variations of the ellipticity and of the azimuth
(inclination) angle of the polarization ellipse. They are  examples
of LDV. The comparison of such dynamical variables for different
coherent states requires that affine parallel transport agrees with
the Fubini-Study metric.

As far as I know the generalized problem of the quantum measurement
of an arbitrary hermitian dynamical variable $\hat{H}=E^{\alpha}
\hat{\lambda}_{\alpha}, \quad \hat{\lambda}_{\alpha} \in AlgSU(N)$
in the operational manner given above was newer done. It is solved
here by the exact analytical diagonalization of an hermitian matrix.
Previously this problem was solved partly in the works
\cite{Onf,Ost,Le7}. Geometrically it looks like embedding ``the
ellipsoid of polarizations" into the iso-space of the adjoint
representation of $SU(N)$. This ellipsoid is associated with the
quadric form $<F|\hat{H}|F>=\sum_1^{N^2-1} E^{\alpha}<F|
\hat{\lambda}_{\alpha}|F>=H_{ab}(E^{\alpha})f^{a*} f^b$ depending on
$N^2-1$ real parameters $E^{\alpha}$. The shape of this ellipsoid
with $N$ main axes is giving by the $2(N-1)$ parameters of the coset
transformations $G/H=SU(N)/S[U(1) \times U(N-1)]=CP(N-1)$ relate to
the $(N-1)$ complex local coordinates of the eigen-state of
$\hat{H}$ in $CP(N-1)$. Its orientation in iso-space $R^{N^2-1}$ is
much more complicated than it was in the case of $R^3$. It is given
by generators of the isotropy group containing $N-1=rank(AlgSU(N))$
independent parameters of ``rotations'' about commutative operators
$\hat{\lambda}_3, \hat{\lambda}_8, \hat{\lambda}_{15},...$ and
$(N-1)(N-2)$ parameters of rotations about non-commutative
operators.  All these $(N-1)^2=(N-1)+(N-1)(N-2)$ gauge angles of the
isotropy group $H=S[U(1) \times U(N-1)]$ of the eigen-state giving
orientation of this ellipsoid in iso-space $R^{N^2-1}$ will be
calculated now during the process of analytical diagonalization of
the hermitian matrix $H_{ab}=<a|\hat{H}|b>$ corresponding to some
dynamical variable $\hat{H}$.

{\bf Stage 1. Reduction of the general Hermitian Matrix to
three-diagonal form.} Let me start from general hermitian $N \times
N$ matrix $\hat{H}$. One should choose some basis in $C^N$. I will
take the standard basis
\begin{eqnarray}
|1>= \left( \matrix {1 \cr 0 \cr 0 \cr . \cr . \cr . \cr 0 }
\right),&|2>= \left( \matrix {0 \cr 1 \cr 0 \cr . \cr . \cr . \cr 0
} \right),...,|N>= \left( \matrix {0 \cr 0 \cr 0 \cr . \cr . \cr .
\cr 1 } \right).
\end{eqnarray}
Now if one choose, say, $N=3$ then the standard Gell-Mann
$\hat{\lambda}$ matrices may be distinguished into the two sets in
respect with for example the state $|1>$ : B-set $\hat{\lambda_1},
\hat{\lambda_2}, \hat{\lambda_4}, \hat{\lambda_5}$ whose exponents
act effectively on the $|1>$, and the H-set $\hat{\lambda_3},
\hat{\lambda_8}, \hat{\lambda_6}, \hat{\lambda_7}$ , whose exponents
that leave $|1>$ intact. For any finite dimension $N$ one may define
the ``I-spin" $(1 \leq I \leq N)$ as an analog of the well known
``T-,U-,V- spins" of the $SU(3)$ theory using the invariant
character of the commutation relations of B-and H-sets
\begin{eqnarray}
[B,B] \in H, \quad [H,H] \in H, \quad [B,H] \in B.
\end{eqnarray}
Let me now to represent our hermitian matrix in following manner
\begin{eqnarray}
\hat{H} = \left( \matrix {0 & H_{01} &... H_{0i} &... H_{0N-1}  \cr
H_{10} & 0 &... 0 &... 0  \cr H_{20} & 0 &... 0 &... 0 \cr . \cr .
\cr . \cr H_{N-1 0} & 0 &... 0 &... 0 } \right)_B \cr + \left(
\matrix {H_{00} & 0 &... 0 &... 0  \cr 0 & H_{11} &... H_{1 i} &...
H_{1 N-1} \cr 0 & H_{21} &... H_{2 i} &... H_{2 N-1}  \cr . \cr .
\cr . \cr 0 & H_{N-1 1} &... H_{N-1 i} &... H_{N-1 N-1} } \right)_H.
\end{eqnarray}
In respect with ket $|1>$ one may to classify the first matrix as
$B-type$ and the second one as a matrix of the $H-type$. I will
apply now the ``squeezing ansatz" \cite{Le4,Le7}.  The first
``squeezing'' unitary matrix is
\begin{equation}
\hat{U}_1= \left( \matrix{1&0&0&.&.&.&0 \cr 0&1&0&.&.&.&0 \cr
.&.&.&.&.&.&. \cr .&.&.&.&.&.&. \cr 0&.&.&.&1&0&0 \cr .&.&.&.&0&\cos
\phi_1&e^{i\psi_1} \sin \phi_1 \cr 0&0&.&.&0&-e^{-i\psi_1} \sin
\phi_1&\cos \phi_1 } \right ).
\end{equation}
The transformation of similarity being applied to our matrix gives
$\hat{H}_1 =\hat{U}_1^+ \hat{H} \hat{U}_1$ with the result for
$\hat{H}_B$ shown for simplicity in the case $N=4$
\begin{eqnarray}
\hat{H}_{B1} = \left( \matrix {0&H_{01}&\tilde{H_{02}}
&\tilde{H_{03}} \cr H_{01}^*&0&0&0\cr \tilde{H_{02}^*} &0&0&0 \cr
\tilde{H_{03}^*} &0&0&0} \right),
\end{eqnarray}
where $\tilde{H_{02}}=H_{02} \cos \phi-H_{03}\sin \phi e^{-i\psi}$
and $\tilde{H_{03}} =H_{02}\sin \phi e^{i\psi} + H_{03} \cos \phi $.
Now one has solve two ``equations of annihilation'' of $\Re (H_{02}
\sin(\phi)e^{i\psi} + H_{03}\cos(\phi))=0$ and $\Im (H_{02}
\sin(\phi)e^{i\psi} + H_{03}\cos(\phi))=0$ in order to eliminate the
last element of the first row and its hermitian conjugate
\cite{Le4,Le7}. This gives us $\phi'_1$ and $\psi'_1$. I will put
$H_{02}=\alpha_{02}+i\beta_{02}$ and
$H_{03}=\alpha_{03}+i\beta_{03}$, then the solution of the
``equations of annihilation'' is as follows:
\begin{eqnarray}
\phi'_1=\arctan
\sqrt{\frac{\alpha_{03}^2+\beta_{03}^2}{\alpha_{02}^2+\beta_{02}^2}},
\cr \psi'_1=\arctan \frac{\alpha_{03} \beta_{02} - \alpha_{02}
\beta_{03}}
{\sqrt{(\alpha_{02}^2+\beta_{02}^2)(\alpha_{03}^2+\beta_{03}^2)}}.
\end{eqnarray}
This transformation acts of course on the second matrix $\hat{H}_H$,
but it easy to see that its structure is intact. The next step is
the similarity transformations given by the matrix with the
diagonally shifted transformation block
\begin{equation}
\hat{U}_2= \left( \matrix{ 1&0&0&.&.&.&0 \cr 0&1&0&.&.&.&0 \cr
.&.&.&.&.&.&. \cr 0&.&.&.&1&0&0 \cr .&.&.&.&0&cos \phi_2&e^{i\psi_2}
sin \phi_2 \cr 0&0&.&.&0&-e^{-i\psi_2}sin \phi_2&cos \phi_2 \cr
0&.&.&.&0&0&1 } \right )
\end{equation}
and the similar evaluation of $\psi'_2$, $\phi'_2$. Generally one
should make $N-2$ steps in order to annulate $N-2$ elements of the
first row. The next step is to represent our transformed $\hat{H}_1
=\hat{U}_1^+ \hat{H} \hat{U}_1$ as follows:
\begin{eqnarray}
\hat{H}_1 = \left( \matrix {0 & \tilde{H}_{01} &0 &... 0 \cr
\tilde{H}_{10} & 0 &\tilde{H}_{12}&... \tilde{H}_{1 N-1} \cr 0 &
\tilde{H}_{21} &0 &... 0 \cr . \cr . \cr . \cr 0 & \tilde{H}_{N-1,1}
&... 0 &... 0 } \right)_B \cr + \left( \matrix {H_{00} & 0 &... 0
&... 0 \cr 0 & \tilde{H}_{11} &...0 &... 0\cr 0 & 0 &...
\tilde{H}_{2 i} &... \tilde{H}_{2 N-1} \cr . \cr . \cr . \cr 0 & 0
&... \tilde{H}_{N-1,i} &... \tilde{H}_{N-1,N-1} } \right)_H.
\end{eqnarray}
Now one should applied the squeezing ansatz in $N-3$ steps for
second row, etc., generally one has $(N-1)(N-2)$ orientation angles.
Thereby we come to the three-diagonal form of the our matrix.

{\bf Stage 2. Diagonalization of the three-diagonal form.} The
eigen-problem for the three-diagonal hermitian matrix is well known,
but I will put it here for the completeness. The eigen- problem
$(\hat{\tilde{H}} - \lambda \hat{E})|\xi>=0$ for the three-diagonal
matrix has the following form
\begin{eqnarray}
\left( \matrix {\tilde{H}_{00}\xi^0&\tilde{H}_{01}\xi^1&0&.&.&.&0
\cr
\tilde{H}_{01}^*\xi^0&\tilde{H}_{11}\xi^1&\tilde{H}_{12}\xi^2&0&.&.&0\cr
0&\tilde{H}_{12}^*\xi^1&\tilde{H}_{22}\xi^2&\tilde{H}_{23}\xi^3&.&.&0
\cr 0&0& \tilde{H}_{23}^*\xi^2&.&.&.&0 \cr 0&0&.&.&.&.&0 \cr
0&0&.&.&.&.&\tilde{H}_{N-1,N-2}\xi^{N-1} \cr
0&0&.&.&.&.&\tilde{H}_{N-1,N-1}\xi^{N-1}} \right) = \left( \matrix
{\lambda \xi^0 \cr \lambda \xi^1 \cr \lambda \xi^2 \cr . \cr . \cr .
\cr \lambda \xi^{N-1}} \right).
\end{eqnarray}
Since $\xi^1=\frac{\lambda -\tilde{H}_{00}}{\tilde{H}_{01}} \xi^0$,
etc., one has the reccurrent relations between all components of the
eigen-vector corresponding to given $\lambda$. Thereby only $N-1$
complex local coordinates
$(\pi^1=\frac{\xi^1}{\xi^0},...,\pi^{N-1}=\frac{\xi^{N-1}}{\xi^0})$
giving the shape of the ellipsoid of polarization have invariant
sense as it was mentioned above.

{\bf Stage 3. The coset ``force'' acting during a measurement} The
real measurement assumes some interaction of the measurement device
and incoming state. If we assume for simplicity that incoming state
is $|1>$ (modulation, etc. are neglected), then all transformations
from $H$-subalgebra will leave it intact. Only the coset unitary
transformations
\begin{eqnarray}
& \hat{T}(\tau,g) = \cr & \left( \matrix{\cos g\tau
&\frac{-p^{1*}}{g} \sin g\tau &\frac{-p^{2*}}{g}\sin g\tau
&.&\frac{-p^{N-1*}}{g}\sin g\tau \cr \frac{p^1}{g} \sin g\tau
&1+[\frac{|p^1|}{g}]^2 (\cos g\tau -1)&[\frac{p^1 p^{2*}}{g}]^2
(\cos g\tau -1)&.&[\frac{p^1 p^{N-1*}}{g}]^2 (\cos g\tau -1) \cr
.&.&.&.&. \cr .&.&.&.&. \cr .&.&.&.&. \cr \frac{p^{N-1}}{g}\sin
g\tau &[\frac{p^{1*} p^{N-1}}{g}]^2 (\cos g\tau
-1)&.&.&1+[\frac{|p^{N-1}|}{g}]^2 (\cos g\tau -1)} \right),
\end{eqnarray}
where $g=\sqrt{|p^1|^2+,...,+|p^{N-1}|^2}$ will effectively to
variate this state dragging it along one of the geodesics in
$CP(N-1)$ \cite{Le4}. This matrix describe the process of the
transition from one pure state to another, in particular between two
eigen-states connected by the geodesic. This means that these
transformations deform the ellipsoid. All possible shapes of these
ellipsoids are distributed along a single geodesic.

Generally, in the dynamical situation this ``stationary'' global
procedure is not applicable and one should go to the local analog of
$\hat{\lambda}$-matrices, i.e. $SU(N)$ generators and related
dynamical variables should be parameterized by the local quantum
states coordinates $(\pi^1,...,\pi^{N-1})$.

\section{Local dynamical variables} The state space ${\cal H}$
with finite action quanta is a stationary construction. We introduce
dynamics {\it by the velocities of the GCS variation} representing
some ``elementary excitations'' (quantum particles). Their dynamics
is specified by the Hamiltonian, giving time variation velocities of
the action quantum numbers in different directions of the tangent
Hilbert space $T_{(\pi^1,...,\pi^{N-1})} CP(N-1)$ which takes the
place of the ordinary linear quantum state space as will be
explained below. The rate of the action variation gives the energy
of the ``particles'' whose expression should be established by some
wave equations.

The local dynamical variables correspond to the internal $SU(N)$
group of the GCS and their breakdown should be expressed now in
terms of the local coordinates $\pi^k$. The Fubini-Study metric
\begin{equation}
G_{ik^*} = [(1+ \sum |\pi^s|^2) \delta_{ik}- \pi^{i^*} \pi^k](1+
\sum |\pi^s|^2)^{-2} \label{FS}
\end{equation}
and the affine connection
\begin{eqnarray}
\Gamma^i_{mn} = \frac{1}{2}G^{ip^*} (\frac{\partial
G_{mp^*}}{\partial \pi^n} + \frac{\partial G_{p^*n}}{\partial
\pi^m}) = -  \frac{\delta^i_m \pi^{n^*} + \delta^i_n \pi^{m^*}}{1+
\sum |\pi^s|^2} \label{Gamma}
\end{eqnarray}
in these coordinates will be used. Hence the internal dynamical
variables and their norms should be state-dependent, i.e. local in
the state space \cite{Le4}. These local dynamical variables realize
a non-linear representation of the unitary global $SU(N)$ group in
the Hilbert state space $C^N$. Namely, $N^2-1$ generators of $G =
SU(N)$ may be divided in accordance with the Cartan decomposition:
$[B,B] \in H, [B,H] \in B, [B,B] \in H$. The $(N-1)^2$ generators
\begin{eqnarray}
\Phi_h^i \frac{\partial}{\partial \pi^i}+c.c. \in H,\quad 1 \le h
\le (N-1)^2
\end{eqnarray}
of the isotropy group $H = U(1)\times U(N-1)$ of the ray (Cartan
sub-algebra) and $2(N-1)$ generators
\begin{eqnarray}
\Phi_b^i \frac{\partial}{\partial \pi^i} + c.c. \in B, \quad 1 \le b
\le 2(N-1)
\end{eqnarray}
are the coset $G/H = SU(N)/S[U(1) \times U(N-1)]$ generators
realizing the breakdown of the $G = SU(N)$ symmetry of the GCS.
Furthermore, the $(N-1)^2$ generators of the Cartan sub-algebra may
be divided into the two sets of operators: $1 \le c \le N-1$ ($N-1$
is the rank of $Alg SU(N)$) Abelian operators, and $1 \le q \le
(N-1)(N-2)$ non-Abelian operators corresponding to the
non-commutative part of the Cartan sub-algebra of the isotropy
(gauge) group. Here $\Phi^i_{\sigma}, \quad 1 \le \sigma \le N^2-1 $
are the coefficient functions of the generators of the non-linear
$SU(N)$ realization. They give the infinitesimal shift of the
$i$-component of the coherent state driven by the $\sigma$-component
of the unitary multipole field $\Omega^{\alpha}$ rotating the
generators of $Alg SU(N)$ and they are defined as follows:
\begin{equation}
\Phi_{\sigma}^i = \lim_{\epsilon \to 0} \epsilon^{-1}
\biggl\{\frac{[\exp(i\epsilon \lambda_{\sigma})]_m^i g^m}{[\exp(i
\epsilon \lambda_{\sigma})]_m^j g^m }-\frac{g^i}{g^j} \biggr\}=
\lim_{\epsilon \to 0} \epsilon^{-1} \{ \pi^i(\epsilon
\lambda_{\sigma}) -\pi^i \},
\end{equation}
\cite{Le4,Le7}. Then the sum of the $N^2-1$ the energies associated
with intensity of deformations of the GCS is represented  by the
local Hamiltonian vector field $\vec{H}$ which is linear in the
partial derivatives $\frac{\partial }{\partial \pi^i} = \frac{1}{2}
(\frac{\partial }{\partial \Re{\pi^i}} - i \frac{\partial }{\partial
\Im{\pi^i}})$ and $\frac{\partial }{\partial \pi^{*i}} = \frac{1}{2}
(\frac{\partial }{\partial \Re{\pi^i}} + i \frac{\partial }{\partial
\Im{\pi^i}})$. In other words it is the tangent vector to $CP(N-1)$
\begin{eqnarray}
\vec{H}=T_c+T_q +V_b = \hbar \Omega^c \Phi_c^i \frac{\partial
}{\partial \pi^i} + \hbar \Omega^q \Phi_q^i \frac{\partial
}{\partial \pi^i} + \hbar \Omega^b \Phi_b^i \frac{\partial
}{\partial \pi^i} + c.c. \label{field}
\end{eqnarray}

In order to express some eigen-vector in the local coordinates, I
put
\begin{eqnarray}
|D_p(\pi^1_{j(p)},...,\pi^{N-1}_{j(p)})>
 =\sum_0^{N-1}
g^a(\pi^1_{j(p)},...,\pi^{N-1}_{j(p)})|\hbar a>,
\end{eqnarray}
where $\sum_{a=0}^{N-1} |g^a|^2= R^2$, and
\begin{eqnarray}
g^0(\pi^1_{j(p)},...,\pi^{N-1}_{j(p)})=\frac{R^2}{\sqrt{R^2+
\sum_{s=1}^{N-1}|\pi^s_{j(p)}|^2}}.
\end{eqnarray}
For $1\leq i\leq N-1$ one has
\begin{eqnarray}
g^i(\pi^1_{j(p)},...,\pi^{N-1}_{j(p)})=\frac{R
\pi^i_{j(p)}}{\sqrt{R^2+\sum_{s=1}^{N-1}|\pi^s_{j(p)}|^2}},
\end{eqnarray}
i.e. $CP(N-1)$ is embedded in the Hilbert space ${\cal{H}}=C^N$.
Hereafter I will suppose $R=1$.

Now we see that all eigen-vectors corresponding to different
eigen-values (even under the degeneration) are applied to different
points $(\pi^1_{j(p)},...,\pi^{N-1}_{j(p)})$ of the $CP(N-1)$.
Nevertheless the eigen-vectors
$|D_p(\pi^1_{j(p)},...,\pi^{N-1}_{j(p)})>$ are mutually orthogonal
in ${\cal{H}}=C^N$ if $\hat{H}$ is hermitian Hamiltonian. Therefore
one has the ``splitting" or delocalization of degenerated
eigen-states in $CP(N-1)$. Thus the local coordinates $\pi^i$  gives
the convenient parametrization of the $SU(N)$ action as one will see
below.

Let me assume that $|G>=\sum_{a=0}^{N-1} g^a|\hbar a>$ is a ``ground
state" of some the least action problem. Then the velocity of the
ground state evolution relative world time is given by the formula
\begin{eqnarray}\label{41}
|H> = \frac{d|G>}{d\tau}=\frac{\partial g^a}{\partial
\pi^i}\frac{d\pi^i}{d\tau}|\hbar
a>=|T_i>\frac{d\pi^i}{d\tau}=H^i|T_i>,
\end{eqnarray}
 is the tangent vector to the evolution curve
$\pi^i=\pi^i(\tau)$, where
\begin{eqnarray}\label{42}
|T_i> = \frac{\partial g^a}{\partial \pi^i}|a\hbar>=T^a_i|\hbar a>,
\end{eqnarray}
and
\begin{eqnarray}
T^0_i=\frac{\partial g^0}{\partial \pi^i} &=&-\frac{1}{2} \frac{
 \pi^{*i}}{\left(\sqrt{\sum_{s=1}^{N-1} |\pi^s|^2+1}\right)^3},\cr
T^m_i=\frac{\partial g^m}{\partial \pi^i}     & = &
\left(\frac{\delta^m_i}{\sqrt{\sum_{s=1}^{N-1} |\pi^s|^2+1}}-
\frac{1}{2} \frac{\pi^m \pi^{*i}} {\left(\sqrt{\sum_{s=1}^{N-1}
|\pi^s|^2+1}\right)^3}\right).
\end{eqnarray}
Then the ``acceleration'' is as follows
\begin{eqnarray}\label{43}
|A> =
\frac{d^2|G>}{d\tau^2}=|g_{ik}>\frac{d\pi^i}{d\tau}\frac{d\pi^k}{d\tau}
+|T_i>\frac{d^2\pi^i}{d\tau^2}=|N_{ik}>\frac{d\pi^i}{d\tau}\frac{d\pi^k}{d\tau}\cr
+(\frac{d^2\pi^s}{d\tau^2}+\Gamma_{ik}^s
\frac{d\pi^i}{d\tau}\frac{d\pi^k}{d\tau})|T_s>,
\end{eqnarray}
where
\begin{eqnarray}\label{44}
|g_{ik}>=\frac{\partial^2 g^a}{\partial \pi^i \partial \pi^k}
|a\hbar>=|N_{ik}>+\Gamma_{ik}^s|T_s>
\end{eqnarray}
and the state
\begin{eqnarray}\label{45}
|N> = N^a|\hbar a>=(\frac{\partial^2 g^a}{\partial \pi^i \partial
\pi^k}-\Gamma_{ik}^s \frac{\partial g^a}{\partial \pi^s})
\frac{d\pi^i}{d\tau}\frac{d\pi^k}{d\tau}|a\hbar>
\end{eqnarray}
is the normal to the ``hypersurface'' of the ground states. Then the
minimization of this ``acceleration'' under the transition from
point $\tau$ to $\tau+d\tau$ may be achieved by the annihilation of
the tangential component
\begin{equation}
(\frac{d^2\pi^s}{d\tau^2}+\Gamma_{ik}^s
\frac{d\pi^i}{d\tau}\frac{d\pi^k}{d\tau})|T_s>=0,
\end{equation}
i.e. under the condition of the affine parallel transport of the
Hamiltonian vector field
\begin{equation}\label{par_tr}
dH^s +\Gamma^s_{ik}H^id\pi^k =0.
\end{equation}

We saw that $SU(N)$ geometry gives the shape and the orientation of
the ellipsoid associated with the ``average" of dynamical variable
given by a quadric form $<F|\hat{D}|F>$. If it is taking ``as given"
it show only primitive eigen-value problem. But if one rises the
question about real operational sense of the quantum measurement of
this dynamical variable or the process of the transition from one
eigen-state to another, one sees that quantum state and dynamical
variable involved in much more complicated relations that it is
given in the orthodox quantum scheme. The simple reason for this is
that the decomposition (representation) of the state vector of a
quantum system strongly depends on the spectrum and eigen-vectors of
its dynamical variable. Overloaded system of the GCS's supplies us
by enough big ``reserve'' of functions but their superposition
should be local and they span a tangent space at any specific point
of $CP(N-1)$ marked by the local coordinates.

The ``probability" may be introduced now by pure geometric way like
$cos^2 \phi $ in tangent state space as follows.

For any two tangent vectors $D_1^i=<D_1|T_i>, D_2^i=<D_2|T_i>$ one
can define the scalar product
\begin{eqnarray}\label{}
(D_1,D_2)=\Re G_{ik^*} D_1^i D_2^{k^*}=\cos \phi_{1,2}
(D_1,D_1)^{1/2} (D_2,D_2)^{1/2}.
\end{eqnarray}
Then the value
\begin{eqnarray}\label{}
P_{1,2}(\pi^1_{j(p)},...,\pi^{N-1}_{j(p)})=\cos^2
\phi_{1,2}=\frac{(D_1,D_2)^2}{(D_1,D_1) (D_2,D_2)}
\end{eqnarray}
may be treated as a relative probability of the appearance of two
states arising during the measurements of two different dynamical
variables $D_1, D_2$ by the variation of the initial GCS
$(\pi^1_{j(p)},...,\pi^{N-1}_{j(p)})$.

Some LDV $\vec{\Psi}=\Psi^i \frac{\partial}{\partial \pi^i} + c.c. $
may be associated with the ``state vector"  $|\Psi> \in \mathcal{H}$
which has tangent components $\Psi^i=<T_i|\Psi>$ in
$T_{\pi}CP(N-1)$. Thus the scalar product
\begin{eqnarray}\label{}
(\Psi,D)=\Re G_{ik^*} \Psi^i D^{k^*}=\cos \phi_{\Psi,D}
(\Psi,\Psi)^{1/2} (D,D)^{1/2}
\end{eqnarray}
gives the local correlation between two LDV's at same GCS. The
cosines of directions
\begin{eqnarray}\label{}
P_{\Psi,i}(\pi^1_{j(p)},...,\pi^{N-1}_{j(p)})=\cos^2
\phi_{\Psi,i}=\frac{(\Psi,D^i)^2}{(\Psi,\Psi) (D^i,D^i)}
\end{eqnarray}
may be identified with ``probabilities" in each tangent direction of
$T_{\pi}CP(N-1)$. The conservation law of ``probability" is given by
the simple identity
\begin{eqnarray}\label{}
\sum_{i=1}^{N-1} P_{\Psi,i}=\sum_{i=1}^{N-1}\cos^2 \phi_{\Psi,i}= 1.
\end{eqnarray}

The notion of the ``probability" is of course justified by our
experience since different kinds of fluctuations prevent the exact
knowledge of any quantum dynamical variable. That is not only
because the uncertainty relation between {\it two} canonically
conjugated dynamical variables puts the limit of accuracy, but
because any real measurement of a {\it single} dynamical variable or
the process of preparation of some state are not absolutely exact.
 It is easy to see from the
relation between the velocity $V^{i}=\frac{d\pi^i}{d\tau} $ in
$CP(N-1)$ and the energy variance $(\Delta H)^2$ through
Aharonov-Anandan relationship $\frac{dS}{d\tau}=\frac{2 \Delta H
}{\hbar}$ \cite{AA90}, where $\Delta H=\sqrt{<\hat H^2>-<\hat H>^2}$
is the uncertainty  of the Hamiltonian $\hat H$. Indeed, the quadric
form in the local coordinates is as follows: $dS^2 =G_{ik^*}d\pi^i
d\pi^{k*}=\frac{4(\Delta H)^2}{\hbar^2}d\tau^2$ and, therefore,
\begin{eqnarray}\label{uncer}
(\Delta H)^2 = \frac{\hbar^2}{4} G_{ik^*} \frac{d\pi^i}{d\tau} \frac
{d\pi^{k*}}{d\tau},
\end{eqnarray}
i.e. velocity $V^{i}$ in $CP(N-1)$ defines the variance of the
Hamiltonian.

But it is not the reason to deny a possibility to know any dynamical
variable with an acceptable accuracy.

\section{Objective Quantum Measurement}
The $CP(N-1)$ manifold takes the place of the ``classical phase
space'' since its points, corresponding to the GCS, are most close
to classical states of motion. Their points may be interpreted as
the ``Schr\"odinger's lump" \cite{Penrose}. It is important that in
this case the ``Schr\"odinger's lump" has the exact mathematical
description and clear physical interpretation: points of $CP(N-1)$
are the axis of the ellipsoid of the action operator $\hat{S}$, i.e.
extremals of the action functional $S[|F>]$. Then the velocities of
variation of these axis correspond to local Hamiltonian or different
local dynamical variables.

Let me assume that GCS described by local coordinates
$(\pi^1,...,\pi^{N-1})$ corresponds to the original lump, and the
coordinates $(\pi^1+\delta \pi^1,...,\pi^{N-1}+\delta \pi^{N-1})$
correspond to the lump displaced due to measurement. I will use a
GCS $(\pi^1_{j(p)},...,\pi^{N-1}_{j(p)})$ of some action operator
$\hat{S}=\hbar A(\hat{\eta^+}\hat{\eta})$ representing physically
distinguishable states. This means that any two points of $CP(N-1)$
define two ellipsoids differ at least by the orientations, if not by
the shape, as it was discussed above. As such, they may be used as
``yes/no'' states of some two-level detector.

Local coordinates of the lump gives the a firm geometric tool for
the description of quantum dynamics during interaction which used
for a measuring process. The question that I would like to raise is
as follows: {\it what ``classical field'', i.e. field in space-time,
corresponds to the transition from the original to the displaced
lump?} In other words I would like to find the measurable physical
manifestation of the lump , which I called the ``field shell", its
space-time shape and its dynamics. The lump's dynamics will be
represented by energy (frequencies) distribution that are not a
priori given, but are defined by some field equations which should
established by means of a new variation problem. Before its
formulation, we wish to introduce differential geometric
construction.

I assume that there is {\it expectation state} $|D>:
\hat{D}|D>=\lambda_p|D>$, associated with ``measuring device'' tuned
for measurement of dynamical variable $\hat{D}$ at some eigen-state
$(\pi^1_{j(p)},...,\pi^{N-1}_{j(p)})$
\begin{eqnarray}
|D>=|D_p(\pi^1_{j(p)},...,\pi^{N-1}_{j(p)})>
 =\sum_{a=0}^{N-1}
g^a(\pi^1_{j(p)},...,\pi^{N-1}_{j(p)})|\hbar a>=\sum_{a=0}^{N-1}
g^a|\hbar a>.
\end{eqnarray}
Hereafter I will omit indexes ${j(p)}$ for a simplicity. Now one
should build the spinor of the ``logical spin 1/2" in the local
basis $(|N>,|\widetilde{D}>)$ for the quantum question in respect
with the measurement of the local dynamical variable $\vec{D}$ at
corresponding GCS which may be marked by the local normal state
\begin{eqnarray}\label{45}
|N> = N^a|\hbar a>=(\frac{\partial^2 g^a}{\partial \pi^i \partial
\pi^k}-\Gamma_{ik}^s \frac{\partial g^a}{\partial \pi^s})
\frac{d\pi^i}{d\tau}\frac{d\pi^k}{d\tau}|\hbar a>.
\end{eqnarray}
Since in general $|D>$ it is not a tangent vector to $CP(N-1)$, the
deviation from GCS during the measurement of $\hat{D}$ will be
represented by tangent vector
\begin{eqnarray}\label{}
|\widetilde{D}>=|D>-<Norm|D>|Norm>=|D>-<N|D>\frac{|N>}{<N|N>}
\end{eqnarray}
defined as the covariant derivative on $CP(N-1)$. This operation is
the orthogonal projector $\hat{Q}$. Indeed,
\begin{eqnarray}
\widetilde{|\widetilde{D}>}= \widetilde{(|D>-<Norm|D>|Norm>)}\cr =
|D>-<Norm|D>|Norm> \cr - <Norm|(|D>-<Norm|D>|Norm>)|Norm> \cr
=|D>-<Norm|D>|Norm> = |\widetilde{D}>.
\end{eqnarray}
This projector $\hat{Q}$ takes the place of dichotomic dynamical
variable (quantum question) for the discrimination of the normal
state $|N>$ (it represents the eigen-state at GCS) the and the
orthogonal tangent state $|\widetilde{D}>$ that represents the
velocity of deviation form GCS. The coherent superposition of two
eigen-vectors of $\hat{Q}$ at the point $(\pi^1,...,\pi^{N-1})$
forms the spinor $\eta$ with the components
\begin{eqnarray}\label{513}
\alpha_{(\pi^1,...,\pi^{N-1})}=\frac{<N|D>}{<N|N>} \cr
\beta_{(\pi^1,...,\pi^{N-1})}=\frac{<\widetilde{D}|D>}
{<\widetilde{D}|\widetilde{D}>}.
\end{eqnarray}
Then from the infinitesimally close GCS
$(\pi^1+\delta^1,...,\pi^{N-1}+\delta^{N-1})$, whose shift is
induced by the interaction used for a measurement, one get a close
spinor $\eta+\delta \eta$ with the components
\begin{eqnarray}\label{514}
\alpha_{(\pi^1+\delta^1,...,\pi^{N-1}+\delta^{N-1})}=\frac{<N'|D>}
{<N'|N'>} \cr \beta_{(\pi^1+\delta^1,...,\pi^{N-1}+\delta^{N-1})}=
\frac{<\widetilde{D}'|D>}{<\widetilde{D}'|\widetilde{D}'>},
\end{eqnarray}
where the basis $(|N'>,|\widetilde{D}'>)$ is the lift of the
parallel transported $(|N>,|\widetilde{D}>)$ from the
infinitesimally close point
$(\pi^1+\delta^1,...,\pi^{N-1}+\delta^{N-1})$ back to the
$(\pi^1,...,\pi^{N-1})$. It is clear that such parallel transport
should be somehow connected with the variation of coefficients
$\Omega^{\alpha}$ in the dynamical space-time.

The covariance relative transition from one GCS to another
\begin{eqnarray}
(\pi^1_{j(p)},...,\pi^{N-1}_{j(p)}) \rightarrow
(\pi^1_{j'(q)},...,\pi^{N-1}_{j'(q)})
\end{eqnarray}
and the covariant differentiation (relative Fubini-Study metric) of
vector fields provides the objective character of the ``quantum
question" $\hat{Q}$ and, hence, the quantum measurement. This serves
as a base for the construction of the dynamical space-time as it
will be shown below.

These two infinitesimally close spinors may be expressed as
functions of $\theta,\phi,\psi,R$ and
$\theta+\epsilon_1,\phi+\epsilon_2,\psi+\epsilon_3,R+\epsilon_4,$
and represented as follows
\begin{eqnarray}\label{s1}
\eta = R \left( \begin {array}{c} \cos \frac{\theta}{2}(\cos
\frac{\phi- \psi}{2} - i\sin \frac{\phi - \psi}{2}) \cr \sin
\frac{\theta}{2} (\cos \frac{\phi+\psi}{2} +i \sin
 \frac{\phi+\psi}{2})  \end {array}
 \right)
 = R\left( \begin {array}{c} C(c-is) \cr S( c_1+is_1)
\end
{array} \right)
\end{eqnarray}
and
\begin{eqnarray}
&\eta+\delta \eta = R\left( \begin {array}{c} C(c-is) \cr S(
c_1+is_1) \end {array} \right) \cr + & R\left( \begin {array}{c}
S(is-c)\epsilon_1-C(s+i c)\epsilon_2+
C(s+ic)\epsilon_3+C(c-is)\frac{\epsilon_4}{R} \cr
 C(c_1+is_1)\epsilon_1+S(ic_1-s_1)\epsilon_2-S(s_1-ic_1)\epsilon_3
+S(c_1+is_1)\frac{\epsilon_4}{R}
\end
{array}
 \right)
\end{eqnarray}
may be connected with infinitesimal ``Lorentz spin transformations
matrix'' \cite{G}
\begin{eqnarray}
L=\left( \begin {array}{cc} 1-\frac{i}{2}\tau ( \omega_3+ia_3 )
&-\frac{i}{2}\tau ( \omega_1+ia_1 -i ( \omega_2+ia_2)) \cr
-\frac{i}{2}\tau
 ( \omega_1+ia_1+i ( \omega_2+ia_2))
 &1-\frac{i}{2}\tau( -\omega_3-ia_3)
\end {array} \right).
\end{eqnarray}
Then accelerations $a_1,a_2,a_3$ and angle velocities $\omega_1,
\omega_2, \omega_3$ may be found in the linear approximation from
the equation
\begin{eqnarray}\label{equ}
\eta+\delta \eta = L \eta
\end{eqnarray}
as functions of the ``logical spin 1/2" spinor components depending
on local coordinates $(\pi^1,...,\pi^{N-1})$.

Hence the infinitesimal Lorentz transformations define small
``space-time'' coordinates variations. It is convenient to take
Lorentz transformations in the following form $ct'=ct+(\vec{x}
\vec{a}) d\tau, \quad \vec{x'}=\vec{x}+ct\vec{a} d\tau
+(\vec{\omega} \times \vec{x}) d\tau$, where I put
$\vec{a}=(a_1/c,a_2/c,a_3/c), \quad
\vec{\omega}=(\omega_1,\omega_2,\omega_3)$ \cite{G} in order to have
for $\tau$ the physical dimension of time. The expression for the
``4-velocity" $ v^{\mu}$ is as follows
\begin{equation}
v^{\mu}=\frac{\delta x^{\mu}}{\delta \tau} = (\vec{x} \vec{a},
ct\vec{a}  +\vec{\omega} \times \vec{x}) .
\end{equation}
The coordinates $x^\mu$ of points in dynamical space-time serve in
fact merely for the parametrization of deformations of the ``field
shell'' arising under its motion according to non-linear field
equations \cite{Le1,Le2}.

\section{Field equations in the dynamical space-time}
Now our aim is to find field equations for $\Omega^{\alpha}$
included in the local Hamiltonian vector field $\vec{H}=\hbar
\Omega^{\alpha}\Phi^i_{\alpha}\frac{\partial}{\partial \pi^i} + c.c
$. These field equations should be established in the dynamical
space-time intrinsically connected with the objective quantum
measurement of the ``elementary lump'' associated with a quantum
particle. At each point $(\pi^1,...,\pi^{N-1})$ of the $CP(N-1)$ one
has an ``expectation value'' of the $\vec{H}$ defined by a measuring
device. But a displaced GCS may by reached along one of the
continuum paths. Therefore the comparison of two vector fields and
their ``expectation values'' at neighboring points requires some
natural rule. The comparison for the same ``particle'' may be
realized by ``field shell'' dynamics along some path in $CP(N-1)$.
For this reason one should have an identification procedure. The
affine parallel transport in $CP(N-1)$ of vector fields is a natural
and the simplest rule for the comparison of corresponding ``field
shells''.

The dynamical space-time coordinates $x^{\mu}$ may be introduced as
the state-dependent quantities, transforming in accordance with the
local Lorentz transformations $x^{\mu} + \delta x^{\mu} =
(\delta^{\mu}_{\nu} + \Lambda^{\mu}_{\nu} \delta \tau )x^{\nu}$
depend on the transformations of local reference frame in $CP(N-1)$
as it was described in the previous paragraph.

Let us discuss now the self-consistent problem
\begin{equation}
v^{\mu} \frac{\partial \Omega^{\alpha}}{\partial x^{\mu} } = -
(\Gamma^m_{mn} \Phi_{\beta}^n+\frac{\partial
\Phi_{\beta}^n}{\partial \pi^n}) \Omega^{\alpha}\Omega^{\beta},
\quad \frac{d\pi^k}{d\tau}= \Phi_{\beta}^k \Omega^{\beta}
\end{equation}
arising under the condition of the affine parallel transport
\begin{eqnarray}
\frac{\delta H^k}{\delta \tau} &= &\hbar \frac{\delta
(\Phi^k_{\alpha} \Omega^{\alpha})}{\delta \tau}=0
\end{eqnarray}
of the Hamiltonian field. I will discuss the simplest case of
$CP(1)$ dynamics when $1\leq \alpha,\beta \leq3,\quad i,k,n=1$. This
system in the case of the spherical symmetry being split into the
real and imaginary parts takes the form
\begin{eqnarray}
\matrix{ (r/c)\omega_t+ct\omega_r=-2\omega \gamma F(u,v), \cr
(r/c)\gamma_t+ct\gamma_r=(\omega^2 - \gamma^2) F(u,v), \cr u_t=\beta
U(u,v,\omega,\gamma), \cr v_t=\beta V(u,v,\omega,\gamma), }
\label{self_sys}
\end{eqnarray}

It is impossible of course to solve this self-consistent problem
analytically even in this simplest case of the two state system, but
it is reasonable to develop a numerical approximation in the
vicinity of the following exact solution. Let me put $\omega=\rho
\cos \psi, \quad \gamma=\rho \sin \psi$, then, assuming for
simplicity that $\omega^2+\gamma^2=\rho^2=constant$, the two first
PDE's may be rewritten as follows:
\begin{equation}
\frac{r}{c}\psi_t+ct\psi_r=F(u,v) \rho \cos \psi.
\end{equation}
The one of the exact solutions of this quasi-linear PDE is
\begin{equation}
\psi_{exact}(t,r)=\arctan \frac{\exp(2c\rho F(u,v)
f(r^2-c^2t^2))(ct+r)^{2F(u,v)}-1}{\exp(2c\rho F(u,v)
f(r^2-c^2t^2))(ct+r)^{2F(u,v)}+1}, \label{ex_sol}
\end{equation}
where $f(r^2-c^2t^2)$ is an arbitrary function of the interval.

In order to keep physical interpretation of these equations I will
find the stationary solution for (58). Let me put $\xi=r-ct$. Then
one will get ordinary differential equation
\begin{equation}
\frac{d\Psi(\xi)}{d \xi} = -F(u,v) \rho \frac{\cos \Psi(\xi)}{\xi}.
\end{equation}
Two solutions
\begin{equation}
\Psi(\xi) =arctan(\frac{\xi^{-2M} e^{-2CM}-1}{\xi^{-2M} e^{-2CM}+1},
\frac{2\xi^{-M} e^{-2CM}}{\xi^{-2M} e^{-2CM}-1}  ),
\end{equation}
where $M=F(u,v) \rho$ are concentrated in the vicinity of the
light-cone looks like solitary waves, see Fig.1.

\begin{figure}
\includegraphics[width=5in]{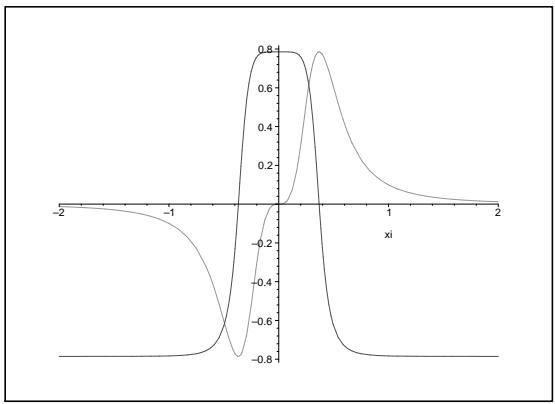}
\caption{Two solutions of (60) in the light-cone vicinity}
\label{fig.1}
\end{figure}

Hence one may treat they as some ``potentials" for the local
coordinates of GCS $(u=\Re \pi^1,v=\Im \pi^1)$. The character of
these solutions should be discussed elsewhere.

\pagebreak {\bf Conclusion}

1. The generalized (in comparison with ``2-level" case \cite{Le6})
the intrinsically geometric scheme of the quantum measurement of an
arbitrary Hermitian ``N-level" dynamical variable has been proposed.
The interaction arose due to the breakdown of $G=SU(N)$ symmetry is
used for such measurement and it is represented by the affine gauge
``field shell" propagated in the dynamical state-dependent
space-time.

2. The concept of ``super-relativity" \cite{Le4} is in fact a
different kind of attempts of ``hybridization" of internal and
space-time symmetries. In distinguish from SUSY where a priori
exists the extended space-time - ``super-space", in my approach the
dynamical space-time arises under ``yes/no" quantum measurement of
$SU(N)$ local dynamical variables.

3. The pure local formulation of quantum theory in $CP(N-1)$ leads
seemingly to the decoherence \cite{Le6}. We may, of course, to make
mentally the concatenation of any two quantum systems living in
direct product of their state spaces. The variation of the one of
them during a measurement may lead formally to some variations in
the second one. Unavoidable fluctuations in our devices may even
confirm predictable correlations. But the introduction of the
state-dependent dynamical space-time evokes a necessity to
reformulate the Bell's inequalities which may lead then to a
different condition for the coincidences.

4. The locality in the quantum phase space $CP(N-1)$ leads to
extended quantum particles - ``field shell" that obey the
quasi-linear PDE \cite{Le1,Le2}. The physical status of their
solutions is the open question. But if they somehow really connected
with ``elementary particles", say, electrons, then the plane waves
of de Broglie should not be literally refer to the state vector of
the electron itself but rather to covector (1-form) realized, say,
by electrons in a periodic cristall lattice. The fact that the
condition for diffraction is in nice agreement with experiments may
be explained that for this agreement it is important only {\it
relative velocity} of electron and the lattice.

ACKNOWLEDGEMENTS

I am grateful to L.P.Horwitz for a lot of interesting discussions
and critical notes, and to A.Schaposhnicov (Tauridian National
University, Simferopol) for prolific discussions in optical
measurements. Further, I thank my wife for her patience during many
years. 

\end{document}